\documentclass[epj]{svjour}
\usepackage{graphics}
\usepackage{amsfonts} 
\begin{document}
\title{Nonadditive entropy: the concept and its use \thanks{To appear in {\it Statistical Power-Law Tails in High Energy Phenomena}, ed. T.S. Biro, Eur. Phys. J. A (2009).}}
\author{Constantino Tsallis
\thanks{tsallis@cbpf.br}
}
%
%
%
\institute{Centro Brasileiro de Pesquisas Fisicas \\
and 
National Institute of Science and Technology for Complex Systems \\       
Xavier Sigaud 150, 22290-180 Rio de Janeiro-RJ, Brazil    \\   
and \\    
Santa Fe Institute, 1399 Hyde Park Road, Santa Fe, NM 87501, USA}
\date{Received: date / Revised version: date}
%
\abstract{
The thermodynamical concept of entropy was introduced by Clausius in 1865 in order to construct the exact differential $dS=\delta Q/T$, where $\delta Q$ is the heat transfer and the absolute temperature $T$ its integrating factor. A few years later, in the period 1872-1877, it was shown by Boltzmann that this quantity can be expressed in terms of the probabilities associated with the microscopic configurations of the system. We refer to this fundamental connection as the Boltzmann-Gibbs (BG) entropy, namely (in its discrete form) $S_{BG}=-k \sum_{i=1}^W p_i \ln p_i$, where $k$ is the Boltzmann constant, and $\{p_i\}$ the probabilities corresponding to the $W$ microscopic configurations (hence $\sum_{i=1}^W p_i=1$). This entropic form, further discussed by Gibbs, von Neumann and Shannon, and constituting the basis of the celebrated BG statistical mechanics, is {\it additive}. Indeed, if we consider a system composed by any two probabilistically independent subsystems $A$ and $B$ (i.e., $p_{ij}^{A+B}=p_i^A p_j^B, \, \forall (i,j))$, we verify that $S_{BG}(A+B)=S_{BG}(A)+S_{BG}(B)$. If a system is constituted by $N$ equal elements which are are either independent or quasi-independent (i.e., not too strongly correlated, in some specific {\it nonlocal} sense), this additivity guarantees $S_{BG}$ to be {\it extensive} in the thermodynamical sense, i.e., that $S_{BG}(N) \propto N$ in the $N>>1$ limit. If, on the contrary, the correlations between the $N$ elements are strong enough, then the extensivity of $S_{BG}$ is lost, being therefore incompatible with classical thermodynamics. In such a case, the many and precious relations described in textbooks of thermodynamics become invalid. Along a line which will be shown to overcome this difficulty, and which consistently enables the generalization of BG statistical mechanics, it was proposed in 1988 the entropy  $S_q=k [1-\sum_{i=1}^W p_i^q]/(q-1)\,(q \in {\cal R}; \, S_1=S_{BG})$. In the context of cybernetics and information theory, this and similar forms have in fact been repeatedly introduced before 1988. The entropic form $S_q$ is, for any $q \neq 1$, {\it nonadditive}. Indeed, for two probabilistically independent subsystems, it satisfies $S_q(A+B)/k=[S_q(A)/k]+[S_q(B)/k]+(1-q)[S_q(A)/k][S_q(B)/k] \ne S_q(A)/k+S_q(B)/k$. This form will turn out to be {\it extensive} for an important class of nonlocal correlations, if $q$ is set equal to a special value different from unity, noted $q_{ent}$ (where $ent$ stands for $entropy$). In other words, for such systems, we verify that $S_{q_{ent}}(N) \propto N \, (N>>1)$, thus legitimating the use of the classical thermodynamical relations. Standard systems, for which $S_{BG}$ is extensive, obviously correspond to $q_{ent}=1$. Quite complex systems exist in the sense that, for them, no value of $q$ exists such that $S_q$ is extensive. Such systems are out of the present scope: they might need forms of entropy different from $S_q$, or perhaps --- more plainly --- they are just not susceptible at all for some sort of thermostatistical approach. Consistently with the results associated with $S_q$, the $q$-generalizations of the Central Limit Theorem and of its extended L\'evy-Gnedenko form have been achieved.  These recent theorems could of course be the cause of the ubiquity of $q$-exponentials, $q$-Gaussians and related mathematical forms in natural, artificial and social systems. All of the above, as well as presently available experimental, observational and computational confirmations --- in high energy physics and elsewhere ---, are briefly reviewed. Finally, we address a confusion which is quite common in the literature, namely referring to distinct physical mechanisms {\it versus} distinct regimes of a single physical mechanism.    
} 
\maketitle
\section{Introduction}
\label{intro}
The concept of entropy $S$, as well as its name, were introduced in thermodynamics by Clausius in 1865 \cite{Clausius1865}. It was done on a purely {\it macroscopic} basis (the possible existence of a {\it microscopic} world was merely speculative at the time, with just a few incipient scientific indications), with the purpose to associate with the heat transfer $\delta Q$ an exact differential. This was indeed achieved through the celebrated relation $dS=\delta Q/T$, where $dS$ is the differential entropy, and the absolute temperature $T$ the integrating factor. 

A decade later, in the period 1872-1877, it was shown by Boltzmann \cite{Boltzmann1872,Boltzmann1877} that this quantity can be expressed in terms of the probabilities associated with the microscopic configurations of the system. We refer to this connection --- one of the deepest ever done in physics --- as the Boltzmann-Gibbs (BG) entropy. In its present (discrete) form, it is written as follows 
\begin{equation}
S_{BG}=-k \sum_{i=1}^W p_i \ln p_i \,, 
\label{BGentropy}
\end{equation}
where $k$ is the Boltzmann constant (or some other convenient value, e.g. $k=1$, in areas outside physics, such as information theory, cybernetics and others), and the probabilities $\{p_i\}$ corresponding to the $W$ microscopic configurations satisfy
\begin{equation}
\sum_{i=1}^W p_i=1 \,.
\end{equation}
This entropic form, further discussed by Gibbs \cite{Gibbs1901}, von Neumann \cite{vonNeumann1927} and Shannon \cite{Shannon1948}, constitutes the basis of the BG statistical mechanics, one of the monuments of contemporary physics. Eq. (\ref{BGentropy}) satisfies a variety of convenient mathematical properties (non-negativity, concavity, expansibility, Lesche-stability, composability, Topsoe factorizability, finite entropy production per unit time satisfying the Pesin identity).  For equal probabilities, i.e., for $p_i=1/W \, (\forall i)$, it attains its maximal value, namely
\begin{equation}
S_{BG}=k \ln W \,. 
\label{BGentropy2}
\end{equation}
In the present context, let us focus on its {\it additivity} property. An entropy $S(\{p_i\})$ is said {\it additive} \cite{Penrose1970} if, for  
a system composed by any two {\it probabilistically independent} subsystems $A$ and $B$ (i.e., satisfying $p_{ij}^{A+B}=p_i^A p_j^B, \, \forall (i,j); \, i=1,2,...,W_A; \, j=1,2,...,W_B)$, we verify 
\begin{equation}
S(A+B)=S(A)+S(B)\,,
\end{equation}
where $S(A+B)\equiv S(\{p_{ij}^{A+B}\})$, $S(A)\equiv S(\{p_{i}^A\})$ and $S(B)\equiv S(\{p_{j}^B\})$. It is straightforward to verify that \\
$S_{BG}(\{p_i\})$ given by Eq. (\ref{BGentropy}) is additive. Due to this property, the BG entropy of any system made of $N$ equal and independent elements satisfies
\begin{equation}
S_{BG}(N)=NS_{BG}(1) \,.
\end{equation}
This fact obviously complies with the classical thermodynamical requirement for the entropy $S$ to be {\it extensive}, i.e., such that
\begin{equation}
\lim_{N \to\infty} \frac{S(N)}{N} < \infty \,.
\label{extensivity}
\end{equation}
Indeed, in such a case, 
\begin{equation}
\lim_{N \to\infty} \frac{S_{BG}(N)}{N}=S_{BG}(1) \le k \ln [W(1)] \,,
\end{equation}
where $W(1)$ is the number, assumed finite, of possible configurations of one element. 

If the system is constituted by $N$ equal elements which are not strictly independent, but quasi-independent instead (i.e., not too strongly correlated, in some  {\it nonlocal} sense to be further clarified later on; typically for a Hamiltonian many-body system whose elements interact through {\it short-range} interactions, or which are weakly quantum -entangled), the additivity of $S_{BG}$ guarantees its extensivity in the thermodynamical sense, i.e., that Eq. (\ref{extensivity}) is satisfied.

If, on the contrary, the correlations between the $N$ elements are strong enough (a feature which might typically occur for nonergodic states, e.g., in Hamiltonian many-body systems with {\it long-range} interactions, or which are strongly quantum-entangled), then the extensivity of $S_{BG}$ might be lost (at least at the level of a large subsystem of a much larger system), being therefore incompatible with classical thermodynamics. In such a case, many of the useful relations described in textbooks of thermodynamics may become invalid. It is precisely this pathological class of systems the one which is addressed within the thermostatistical theory usually referred to as {\it nonextensive statistical mechanics} \cite{Tsallis1988,CuradoTsallis1991,TsallisMendesPlastino1998}, described in the next Section.

\section{Nonadditive entropy and nonextensive statistical mechanics}
\label{sec:1}

\subsection{Nonadditive entropy $S_q$}

As an attempt to generalize BG statistical mechanics, and possibly provide a frame for handling some of the above mentioned pathological systems, it was postulated in 1988 \cite{Tsallis1988} the following entropy:
\begin{equation}
S_q=k \frac{1-\sum_{i=1}^W p_i^q}{q-1} \,\,\, (q \in {\mathbb{R}}; \,S_1=S_{BG}), 
\label{qentropy}
\end{equation}
A simple manner to obtain $S_1=S_{BG}$ is through the use of $p_i^{q-1}=e^{(q-1) \ln p_i}\sim 1+(q-1)\ln p_i$. If $q<0$, the sum must be done only over configurations which have nonzero probability to occur. 
Entropy (\ref{qentropy}) can be conveniently rewritten in the following alternative forms:
\begin{eqnarray}
S_q&=&k \sum_{i=1}^W p_i \ln_q (1/p_i) \\
   &=&-k \sum_{i=1}^W p_i^q \ln_q p_i \\
   &=&-k \sum_{i=1}^W p_i \ln_{2-q} p_i \,,
\label{qentropy2}
\end{eqnarray}
where the {\it $q$-logarithmic function} is defined as follows:
\begin{equation}
\ln_q z \equiv \frac{z^{1-q}-1}{1-q}\,\,(q \in \mathbb{R}; \,z \ge0; \,\ln_1 z=\ln z)\,. 
\label{qlog}
\end{equation}
$S_q$ attains its extremum (maximum for $q>0$, and minimum for $q<0$) for equal probabilities, and its value is given by
\begin{equation}
S_q=k\ln_q W \,.
\label{equalprob}
\end{equation}
It can be shown to satisfy, for independent subsystems,
\begin{equation}
\frac{S_q(A+B)}{k}=\frac{S_q(A)}{k}+\frac{S_q(B)}{k}+(1-q)\frac{S_q(A)}{k} \frac{S_q(B)}{k} \,.
\label{nonadditive}
\end{equation} 
Therefore, this entropy is generically {\it nonadditive}. It satisfies,  nevertheless, most other properties (mentioned previously) of the entropy $S_{BG}$. In other words, it constitutes a sort of minimalistic generalization of $S_{BG}$. From Eq. (\ref{nonadditive}) we obtain
\begin{equation}
S_q(A+B)=S_q(A)+S_q(B)+\frac{1-q}{k}\,S_q(A) \, S_q(B) \,,
\label{nonadditive2}
\end{equation} 
which exhibits the equivalence between $(q-1) \to 0$ and $k \to\infty$. Since for stationary states (e.g., thermal equilibrium), $k$ appears multiplicatively accompanied by the temperature $T$ (i.e., in the form $kT$), $k\to\infty$ turns out to be equivalent to $T\to\infty$. We may consider that it is here where the fact emerges that, at the $T\to\infty$ limit, {\it all} the microcanonical, canonical, grand-canonical ensembles of classical or quantum (Fermi-Dirac, Bose-Einstein, Gentile parastatistics) statistics, $q$-statistics (as will become evident later on)), coincide, and coincide with the hypothesis of equal probabilities for an isolated system.   

Eq. (\ref{nonadditive}) can be generalized in the presence of arbitrary correlations between two sysbsystems $A$ and $B$ of a given system. It becomes \cite{Abe2000}
\begin{eqnarray}
\frac{S_q(A+B)}{k}=\frac{S_q(A)}{k}+\frac{S_q(B|A)}{k}+(1-q)\frac{S_q(A)}{k} \frac{S_q(B|A)}{k} \nonumber \\
   \hspace{-1.5cm}=\frac{S_q(A|B)}{k}+\frac{S_q(B)}{k}+(1-q)\frac{S_q(A|B)}{k} \frac{S_q(B)}{k}\;,\;\;\;\;\;  
\label{nonadditive2}
\end{eqnarray}
where $S_q(A+B)$ is to be calculated with the {\it joint} probabilities $\bigl\{p_{ij}^{A+B}\bigr\}$, $S_q(A)$ with the {\it marginal} probabilities $ \bigl\{p_{i}^{A}\bigr\} \equiv \bigl\{\sum_{j=1}^{W_B} p_{ij}^{A+B}\bigr\}$ (analogously for $S_q(B)$), and \\
$S_q(A|B)$ with the {\it conditional} probabilities $\bigl\{p_{ij}^{A+B}/p_j^B\bigr\}$ (analogously for $S_q(B|A)$): see \cite{Abe2000} for full details. Eq. (\ref{nonadditive2}) straightforwardly recovers Eq. (\ref{nonadditive}) if $A$ and $B$ are independent, hence $S_q(A|B)=S_q(A)$ and $S_q(B|A)=S_q(B)$. 

It is precisely this nonadditivity the property which enables thermodynamical extensivity. More precisely, if both $A$ and $B$ are very large (i.e., $N_A >>1$ and $N_B >>1$), then a value of $q$, noted $q_{ent}$, might exist for which $S_{q_{ent}}(A+B) \sim S_{q_{ent}}(A)+S_{q_{ent}}(B)$. In other words, if a system has $N>>1$ equal elements, it becomes possible that a special value of $q$ exists such that generically $0< \lim_{N \to\infty} 
\bigl[ S_{q_{ent}}(N)/N \bigr] <\infty$. 

This interesting feature can be easily illustrated in the case of equal probabilities, for which Eq. (\ref{equalprob}) holds. If $W(N) \sim C\mu^N$ (with $C>0$ and $\mu>1$), then $q_{ent}=1$, i.e., $S_{BG}(N) \propto N \;\;(N>>1)$. But if strong correlations forbid many (typically most) microscopic configurations to occur, then it might happen that the number $W_{eff}(N)$ of {\it effective} (or {\it admissible}) configurations satisfies $W_{eff}(N)<<W(N)$. If we have, in particular, $W_{eff} \sim DN^\rho$ (with $D>0$ and $\rho \in \mathbb{R}$), then 
\begin{equation}
q_{ent}=1 - \frac{1}{\rho}   \,.
\end{equation}
This type of highly restricted phase space may occur in various systems, as has been numerically or analytically illustrated in various examples. Let us briefly mention here two of them that are analytically tractable, namely an abstract probabilistic one and a physical one. 

The probabilistic model consists in $N$ correlated distinguishable binary variables \cite{TsallisGellMannSato2005}. The probabilities of the $2^N$ states vanish excepting for $\; \sim (d+1)N$ of them (which can be seen, in the classical Pascal-like triangular representation, as a $N$-long ``strip" whose width is $(d+1)$, $d$ being non-negative). This model asymptotically satisfies probabilistic scale-invariance ({\it Leibnitz triangle rule}) in the limit $N\to\infty$.  It can be verified that
\begin{equation}
q_{ent}=1 - \frac{1}{d}   \,.
\end{equation}

The physical model corresponds to a long ring of $N$ $1/2$ spins with ferromagnetic first-neighbor interactions at zero temperature. The interactions are of the anisotropic $XY$ ones in the presence of a transverse external magnetic field (i.e., along the $Z$ direction) at its critical value. Two well known universality classes are contained within such a system, namely the Ising universality class (corresponding to a {\it central charge} $c=1/2$), and the isotropic $XY$ universality class (corresponding to a central charge $c=1$). We consider a block of $L$ successive spins among those $N$ spins, and address the entropy $S_q(L)$ of the $N \to\infty$ quantum system. More precisely, we are interested in
\begin{equation}
S_q(L) = k \frac{1-Tr (\rho_L)^q}{q-1} \,,
\end{equation}
where $\rho_L \equiv \lim_{N\to\infty} Tr_{\{N-L\}} \rho(N)$, $\rho(N)$ being the density matrix associated with the system of $N$ spins, and where we have traced over all but the $L$ successive spins. We define, in this case, $q_{ent}$ as the value of $q$ for which the {\it block entropy} $S_q(L)$ is extensive, i.e., such that $S_{q_{ent}}(L) \propto L$. Such value does exist \cite{CarusoTsallis2008}, and it is given by $q_{ent}= \sqrt{37}-6 \simeq 0.0828$ for $c=1/2$, and 
$q_{ent}= \sqrt{10}-3 \simeq 0.1623$ for $c=1$. By using a recent result within conformal quantum field theory \cite{CalabreseCardy2004}, these two values for $q_{ent}$  can be generalized for the entire class of $(1+1)$-dimensional models characterized by a generic central charge $c$. It is obtained \cite{CarusoTsallis2008}
\begin{equation}
q_{ent}=\frac{\sqrt{9+c^2}-3}{c}   \,.
\end{equation}
As we see, $q_{ent}$ monotonically increases from zero to unity (the BG value!) when $c$ increases from zero to infinity. For $c=4$, one obtains $q=1/2$, which has already emerged in general relativistic problems \cite{OliveiraSoaresTonini2008}; the possible connection, if there is one, remains however without explanation at the present time. Also, for $c=26$, which corresponds to string theory \cite{GinspargMoore1993}, we obtain $q_{ent} \simeq 0.8913$. Finally, it also remains presently without explanation the reason for which $c\to\infty$ leads to the BG result, i.e., $q_{ent}=1$. 
  
The previous system is a fermionic $d=1$ one, for which it is known that $S_{BG}(L) \propto \ln L$. Results, though only numerical, also exist for a bosonic $d=2$ system \cite{CarusoTsallis2008}. Once again extensivity only occurs for $q_{ent}<1$, whereas $S_{BG}(L) \propto L$. Finally it is known that, for the black hole, the BG entropy is proportional to the area $L^2$, instead of being proportional to the volume $L^3$ (see, for instance, \cite{Maddox1993}); in fact, more generally, it is known, for $d$-dimensional bosonic systems, the so-called {\it area law}, i.e., the fact that $S_{BG}(L) \propto L ^{d-1}$, which obviously violates classical thermodynamics. All these anomalies are believed to be a consequence of strongly nonlocal quantum entanglement. In fact all of the above results can be unified through the following conjectural expression, $\forall d$:
\begin{equation}
S_{BG}(L) \propto \ln_{2-d} L \equiv \frac{L^{d-1}-1}{d-1} \ne L^d \propto N \,.
\end{equation}
In all these cases, as for the $d=1$ and the $d=2$ above described examples, it might well exist a value of $q_{ent}<1$ such that thermodynamic extensivity is ensured, i.e., $S_{q_{ent}}(L) \propto L^d \propto N$. This would of course mean that Clausius entropy should be, for this class of anomalous (sub)systems, identified with $S_{q_{ent}}$ and not with $S_{BG}$. The plausible conjectural scenario is  summarized in Table 1. 

\begin{table}
\caption{QSS stands for {\it quasi-stationary state} (\cite{PluchinoRapisardaTsallis2007,PluchinoRapisardaTsallis2008} and references therein).}
\label{tab:1}       
\begin{tabular}{lll}
\hline\noalign{\smallskip}
SYSTEM & ENTROPY $S_{BG}$ & ENTROPY $S_q$ ($q<1$)  \\
   & (additive) &  (nonadditive) \\
\noalign{\smallskip}\hline\noalign{\smallskip}
Short-range \\interactions, \\weakly \\entangled \\blocks, etc & $\;\;$  {\bf EXTENSIVE} & NONEXTENSIVE \\
\noalign{\smallskip}
Long-range \\interactions \\(QSS), \\ strongly \\entangled \\ blocks, etc& NONEXTENSIVE & $\;\;$ {\bf EXTENSIVE} \\
\noalign{\smallskip}\hline
\end{tabular}
\end{table}

\subsection{Nonextensive statistical mechanics}
\label{sec:2}

Since $S_q$ generalizes $S_{BG}$ and maintains most of its  mathematically convenient properties (e.g., concavity, Lesche stability, among others), it is quite natural to attempt the $q$-generalization of BG statistical mechanics itself. This extended theory is usually referred to in the literature as {\it nonextensive statistical mechanics}.  It was first proposed in 1988 \cite{Tsallis1988}, and later on connected to thermodynamics \cite{CuradoTsallis1991,TsallisMendesPlastino1998}. It has since then received a considerable amount of applications and verifications in natural, artificial and social systems \cite{bibliography,GellMannTsallis2004,BoonTsallis2005,Tsallis2008,Tsallis2009}. Some of its predictions have been experimentally and observationally checked in systems such as the motion of {\it Hydra viridissima} \cite{UpadhyayaRieuGlazierSawada2001} and cells \cite{ThurnerWickHanelSedivyHuber2003,DiambraCintraSchubertCosta2005,DiambraCintraChenSchubertCosta2006}, defect turbulence \cite{DanielsBeckBodenschatz2004}, solar wind \cite{BurlagaVinas2005,BurlagaVinasNessAcuna2006,BurlagaVinasWang2007,BurlagaVinas2007}, cold atoms in optical dissipative lattices \cite{DouglasBergaminiRenzoni2006}, dusty plasma \cite{LiuGoree2008}, silo drainage \cite{ArevaloGarcimartinMaza2007a,ArevaloGarcimartinMaza2007b}, high-energy physics (see, e.g., \cite{BediagaCuradoMiranda2000,Beck2000,TsallisAnjosBorges2003,ConroyMiller2008,presentvolume}). They have also been checked  analytically and computationally in various nonlinear dynamical problems such as the edge of chaos of simple unimodal dissipative maps \cite{TsallisPlastinoZheng1997,TirnakliTsallisLyra1999,Tirnakli2000,LatoraBarangerRapisardaTsallis2000,BaldovinRobledo2002,BaldovinRobledo2004,MayoralRobledo2004,Robledo2005,MayoralRobledo2005,TirnakliBeckTsallis2007,TirnakliTsallisBeck2008}, and long-range-interacting many-body Hamiltonian systems \cite{AnteneodoTsallis1998,LatoraRapisardaTsallis2001,PluchinoRapisardaTsallis2007,PluchinoRapisardaTsallis2008}; also, various  applications to the so-called scale-free networks are available \cite{SoaresTsallisMarizSilva2005,ThurnerTsallis2005,Thurner2005,WhiteKejzarTsallisFarmerWhite2006,MenesesCunhaSoaresSilva2006,Hasegawa2006,LiWangNivanenLeMehaute2006,ThurnerKyriakopoulosTsallis2007,Tsallis2008b}.  In one way or another, most if not all of these systems appear to share {\it slow} (power-law rather than exponential) sensitivity to the initial conditions. In other words, for classical systems, at the level of first principles, BG statistical mechanical concepts are legitimate and fruitful when the system exhibits a {\it positive} maximal Lyapunov exponent (corresponding essentially to Boltzmann's {\it molecular chaos hypothesis}), whereas {\it vanishing} maximal Lyapunov exponent appears to be necessary (although probably not sufficient) for the applicability of the nonextensive statistical mechanical concepts. The mechanisms that typically yield $q$-statistics involve, at the mesoscopic level, non-Markovian processes \cite{FuentesCaceres2008}, multiplicative noise \cite{AnteneodoTsallis2003}, nonlinear Fokker-Planck equations \cite{PlastinoPlastino1995,TsallisBukman1996,SchwammleCuradoNobre2007,SchwammleNobreCurado2008}, and similar ones.   

Let us now briefly review, within the present theory, two important stationary-state distributions, namely the $q$-generalization of the celebrated BG weight (discrete case), and the $q$-generalization of the Gaussian distribution (continuous case).

To generalize the BG factor for the canonical ensemble (i.e., a system in a stationary state due to its contact with a ``thermostat") we follow \cite{TsallisMendesPlastino1998}. We must extremize $S_q$ as given by Eq. (\ref{qentropy}) with the constraints
\begin{equation}
\sum_{i=1}^W p_i=1 \,,
\label{normalization}
\end{equation}
and
\begin{equation}
\sum_{i=1}^W E_iP_i^{(q)}=U_q \,,
\label{qinternalenergy}
\end{equation}
where the {\it escort distribution} $\{P_i^{(q)}\}$ is defined through
\begin{equation}
P_i^{(q)} \equiv \frac{p_i^q}{\sum_{j=1}^W p_j^q} \;\;, \;\;p_i = \frac{[P_i^{(q)}]^{1/q}}{\sum_{j=1}^W [P_j^{(q)}]^{1/q}} \,,
\end{equation}
and $U_q$ is a {\it finite} quantity characterizing the {\it width} of the energy distribution $\{p_i\}$; $\{E_i\}$ are the eigenvalues of the system Hamiltonian (with the chosen boundary conditions). Notice, by the way, that constraint (\ref{normalization}) can equivalently be written as $\sum_{i=1}^W P_i^{(q)}=1$. 

Through the entropic optimization procedure, we obtain straighforwardly 

\begin{equation}
p_i =\frac{e_{q}^{- \beta_q (E_i-U_q)}}{\bar{Z}_q} \,,  
\label{qprobability}
\end{equation}
with
\begin{equation}
\beta_q \equiv \frac{\beta}{\sum_{j=1}^W p_j^q} \,,
\label{3.6.9d}
\end{equation}
and
\begin{equation}
\bar{Z}_q \equiv \sum_i^W e_{q}^{- \beta_q (E_i-U_q)} \,,
\label{3.6.9e}
\end{equation}
with the {\it $q$-exponential function} (inverse of the previously defined $q$-logarithm)  $e_q^z \equiv [1+(1-q)z]^{1/(1-q)}$ if $1+(1-q)z>0$, and zero otherwise ($e_1^z=e^z$), 
$\beta$ being the Lagrange parameter associated with the constraint (\ref{qinternalenergy}). Eq. (\ref{qprobability}) makes explicit that the probability distribution is, for fixed $\beta_q$, invariant with regard to the arbitrary choice of the zero of energies. 
The stationary state (or (meta)equilibrium) distribution (\ref{qprobability}) can be rewritten as follows: 
\begin{equation}
p_i= \frac{ e_q^{-\beta_q^\prime E_i}     }{Z_q^\prime} \;,
\label{qprobabilityconvenient}
\end{equation}
with
\begin{equation}
Z_q^\prime \equiv \sum_{j=1}^W e_q^{-\beta_q^\prime E_j} \;,
\label{3.6.9ff}
\end{equation}
and
\begin{equation}
\beta_q^\prime \equiv \frac{\beta_q}{1+(1-q) \beta_qU_q}\;.
\end{equation}
The form (\ref{qprobabilityconvenient}) is particularly convenient for many applications where comparison with experimental or computational data is involved. Also, it makes clear that $p_i$ asymptotically decays like $1/E_i^{1/(q-1)}$ for $q>1$, and has a cutoff for $q<1$, instead of the familiar exponential decay with $E_i$ for $q=1$.

The connection to thermodynamics is established in what follows. It can be proved that
\begin{equation}
\frac{1}{T}=\frac{\partial S_q}{\partial U_q}\;,
\end{equation}
with $T \equiv 1/(k\beta)$. Also we can prove, for the free energy,  
\begin{equation}
F_q \equiv U_q-TS_q= -\frac{1}{\beta} \ln_q Z_q\;,
\end{equation}
where
\begin{equation}
\ln_q Z_q = \ln_q {\bar Z}_q - \beta U_q\;.
\end{equation}
This relation takes into account the trivial fact that, in contrast with what is usually done in BG statistics, the energies $\{E_i\}$ are here referred to $U_q$ in Eq. (\ref{qprobability}). It can also be proved
\begin{equation}
U_q =-\frac{\partial}{\partial \beta} \ln_q Z_q \;,
\end{equation}
as well as relations such as
\begin{equation}
C_q \equiv T\frac{\partial S_q}{\partial T} = \frac{\partial U_q}{\partial T} = -T \frac{\partial^2 F_q}{\partial T^2} \;.
\end{equation}
In fact the entire Legendre transformation structure of thermodynamics is $q$-invariant, which is both remarkable and welcome.

As a final remark, let us stress an interesting feature concerning $q>1$ (power-law decay for $p_i$). Let us assume that the energy spectrum has a (quasi-continuous) density state g(E). The normalization condition (\ref{normalization}) implies that $\int dE \,g(E)p(E)$ is finite. Since, for $q>1$, $p(E)$ decays as $1/E^{1/(q-1)}$, it follows that $g(E)/E^{1/(q-1)}$ must be integrable at infinity. This determines the maximal value of $q$ which is mathematically admissible in the present theory (for instance, if $g(E)$ is constant, then $q<2$ must be satisfied). Let us now focus on the other constraint, namely Eq. (\ref{qinternalenergy}). We immediately see that $E g(E) [p(E)]^q$ must also be integrable at infinity, i.e., $E g(E)/E^{q/(q-1)}$ must be integrable, which implies the {\it same limit for $q$ as before!} So, for instance, if $g(E)$ is a constant, both the normalization and the energy constraints are finite for $q<2$. For $q \ge 2$ the entire theory becomes mathematically inadmissible.  

Let us address now the continuous case which generalizes the Gaussian distribution. We want to extremize
\begin{equation}
S_q=k\frac{1-\int dx \,[p(x)]^q}{q-1}
\end{equation}
with the constraints
\begin{equation}
\int dx \,p(x=1 \,,
\end{equation}
and
\begin{equation}
\int dx \,x^2 P^{(q)}(x)= \sigma^2 \,, 
\label{sigma2}
\end{equation}
where
\begin{equation}
P^{(q)}(x)\equiv \frac{[p(x)]^q}{\int dx \, [p(x)]^q} \,,
\end{equation}
$\sigma^2$ being a fixed (positive) quantity. We obtain
\begin{equation}
p(x)= \frac{e_q^{-\beta x^2}}{\int dx \, e_q^{-\beta x^2}} \,,
\end{equation}
where $\beta >0$ can be determined by using constraint (\ref{sigma2}). The entire theory is valid for $q<3$, above which both the normalization and the $q$-variance (\ref{sigma2}) diverge. For $q \ge 1$ the distribution is defined for all values of $x$; for $q<1$ it has a finite support. For $q<5/3$ the standard variance $\int dx \, x^2 p(x)$ is finite; for $q \ge 5/3$ it diverges. See \cite{ThistletonMarshNelsonTsallis2007} for a numerical comparison between variance and $q$-variance, which exhibits the considerable convenience of the latter.

\subsection{$q$-generalized central limit theorems}

We focus here on the $q$-generalization of the Central Limit Theorem (CLT). Let us remind what the standard CLT states essentially. Consider $N$  equal and {\it independent} random variables $\{x_j\}$, $j=1,2,3,...,N$, (in fact, quasi- independent random variables, in some specific sense that is out of the present brief review, are also admissible) whose mean value and variance are {\it finite}. For simplicity we assume that the associated distributions $p(x)$ (one and the same, $\forall j$) are symmetric (hence the mean value vanishes). We are interested in the sum $X_N \equiv \sum_{j=1}^N x_j$ (proportional to the {\it arithmetic average}, which is the quantity that is typically measured in all generic experiments), and in its correponding distribution $P(X_N)$. The CLT states that, after appropriate centering and scaling (with $\sqrt{N}$), $P(X_N)$ converges, in the limit $N\to\infty$, onto a Gaussian distribution (whose variance coincides in fact with that of $p(x)$). 

Another important theorem, sometimes referred to as the L\'evy-Gnedenko theorem, exists for the case when $p(x)$ has a {\it divergent} variance (e.g., the Cauchy-Lorentz distribution). The random variables are still assumed {\it independent} (or quasi-independent, as previously mentioned). If the asymptotic behavior of $p(x)$ decays like $a/|x|^{1+\alpha}$ ($a>0$ and $0<\alpha<2$), then, in the limit $N\to\infty$, the attractor is the L\'evy distribution $L_\alpha(x)$ (also called $\alpha$-stable distribution) which, after appropriate centering and scaling (with $N^{1/\alpha}$), decays like $p(x)$ itself, i.e., like $a/|x|^{1+\alpha}$. 

These two important attractors (in the space of probabilities) emerge whenever the $N$ variables are independent or quasi-independent. {\it What happens if they are not, i.e. if they are strongly correlated?} The nature of the possible attractor depends on the class of correlations. {\it What happens then if this correlation is of the $q$-independent class?} The precise definition of $q$-independence ($1$-independence being just plain  independence), is given in \cite{UmarovTsallisSteinberg2008} (see also \cite{TsallisQueiros2007,QueirosTsallis2007} as well as \cite{UmarovTsallisGellMannSteinberg2008}). It is essentially associated with the strong correlations which emerge in the presence of probabilistic {\it scale-invariance} (Leibnitz triangle rule, in the case of $N$ binary random variables; see \cite{RodriguezSchwammleTsallis2008} and references therein) \footnote{The specific $q$-independent correlation is based on the $q$-Fourier transform \cite{UmarovTsallisSteinberg2008}, whose inverse is in fact more complex than that of the standard Fourier transform (corresponding to $q=1$). Indeed, it involves some subtleties presently under study.}.  

If the $N$ variables are $q$-independent then the attractors are $q$-Gaussians with $q>1$ if an appropriately $q$-generalized variance of $p(x)$ is {\it finite} \cite{UmarovTsallisSteinberg2008} (see \cite{NelsonUmarov2008} for the $q<1$ case), and the so-called $(q,\alpha)$-stable distributions if this variance {\it diverges} (and $p(x)$ asymptotically decays as a power-law) \cite{UmarovTsallisGellMannSteinberg2008}. For $q$-independent random variables, we may summarize the situation as follows: if the appropriately $q$-generalized variance is finite, the attractor is a $q$-Gaussian (Gaussian if the variables are independent, and $q$-Gaussians with $q \ne 1$ otherwise); if it diverges, the attractor is a $(q,\alpha)$-distribution (L\'evy distributions if the variables are independent, and $(q,\alpha)$-distribution with $q \ne 1$ otherwise) \footnote{The $(q,\alpha)$-distributions have, for $q>1$ and $\alpha <2$, two distinct power-law regimes, namely the {\it intermediate} and the {\it distant} ones (see details in \cite{TsallisQueiros2007,QueirosTsallis2007}). The intermediate power-law approaches $q$-Gaussians in the $\alpha \to 2$ limit. The distant one approaches L\'evy distributions in the $q\to 1$ limit. 
}. 

Putting all this together, we may alert the reader about an error commonly appearing in the literature. If (computational, or experimental, or observational) data exhibit an {\it asymptotic} power-law behavior, many authors claim having shown the emergence of a L\'evy distribution. This is clearly incorrect: it could be a L\'evy distribution, or a $q$-Gaussian (which only concides with a L\'evy distribution for $q=2$, in which case it corresponds to the $\alpha=1$ L\'evy distribution, i.e., the Cauchy-Lorentz distribution), or a $(q,\alpha)$-stable distribution, or even none of them! Only the study of the {\it central part} of the empiric distribution (i.e., the non-asymptotic region) could in principle discriminate between all these possibilities. For example, L\'evy distributions with $1 < \alpha <2$ always present an inflection point if represented in log-log scales; $q$-Gaussians do {\it not} have such inflection point in the same representation (see \cite{TsallisQueiros2007,QueirosTsallis2007} for illustrations of this fact).  

The ubiquity of Gaussians in nature is generally believed to be a consequence of the standard CLT. Analogously, one expects the ubiquity of $q$-Gaussian distributions in asymptotically scale-invariant natural, artificial and social systems to be a consequence of the presently described $q$-generalized CLT. Two such systems are under active investigation nowadays, namely unimodal maps, and long-range classical Hamiltonians. 

Simple unimodal dissipative maps $x_{t+1}=f(x_t)$ (like the logistic and similar ones) exhibit, for control parameter values such that the Lyapunov exponent is positive (i.e., strong chaos), sums of many successive iterates which approach Gaussians. If the control parameter is, instead, close to the edge of chaos (where the Lyapunov exponent is zero), $q$-Gaussians appear to be gradually approached by the sums of many successive iterates:  see details in \cite{TirnakliBeckTsallis2007,TirnakliTsallisBeck2008,RuizTsallis2008}.   

Long-range-interacting many-body classical Hamiltonians such as the HMF one \cite{AntoniRuffo1995} are known to present thermostatistical anomalies. Indeed, if isolated (i.e., micro-canonical ensemble), they can present, for a range of values for the total energy and some classes of initial conditions (e.g., the so-called water-bag and double-water-bag ones) a {\it negative} specific heat. Furthermore, they exhibit a maximal Lyapunov exponent which vanishes in the $N \to \infty$ limit \cite{AnteneodoTsallis1998}, a fact which seemingly breaks down the validity of Boltzmann's molecular chaos hypothesis. In such a system, various other anomalies have been exhibited such as aging, spin-glass nature and non-Maxwellian distribution of velocities (see \cite{RapisardaPluchino2005} and references therein). This state is usually referred in the literature as the QSS ({\it quasi-stationary state}) or metastable one. It has been recently exhibited that the system is nonergodic in the QSS, its ensemble-averaged and its time-averaged distribution of velocities being {\it both} non-Maxwellian and different among them. Its time-averaged one, which is the typically observed in experiments, appears to approach a $q$-Gaussian: see details in \cite{PluchinoRapisardaTsallis2007,PluchinoRapisardaTsallis2008} (see also \cite{FigueiredoRochaAmato2008,PluchinoRapisardaTsallis2008b}).  
  
\begin{figure}
\hspace{-2.5cm}
\resizebox{2\columnwidth}{!}{%
  \includegraphics{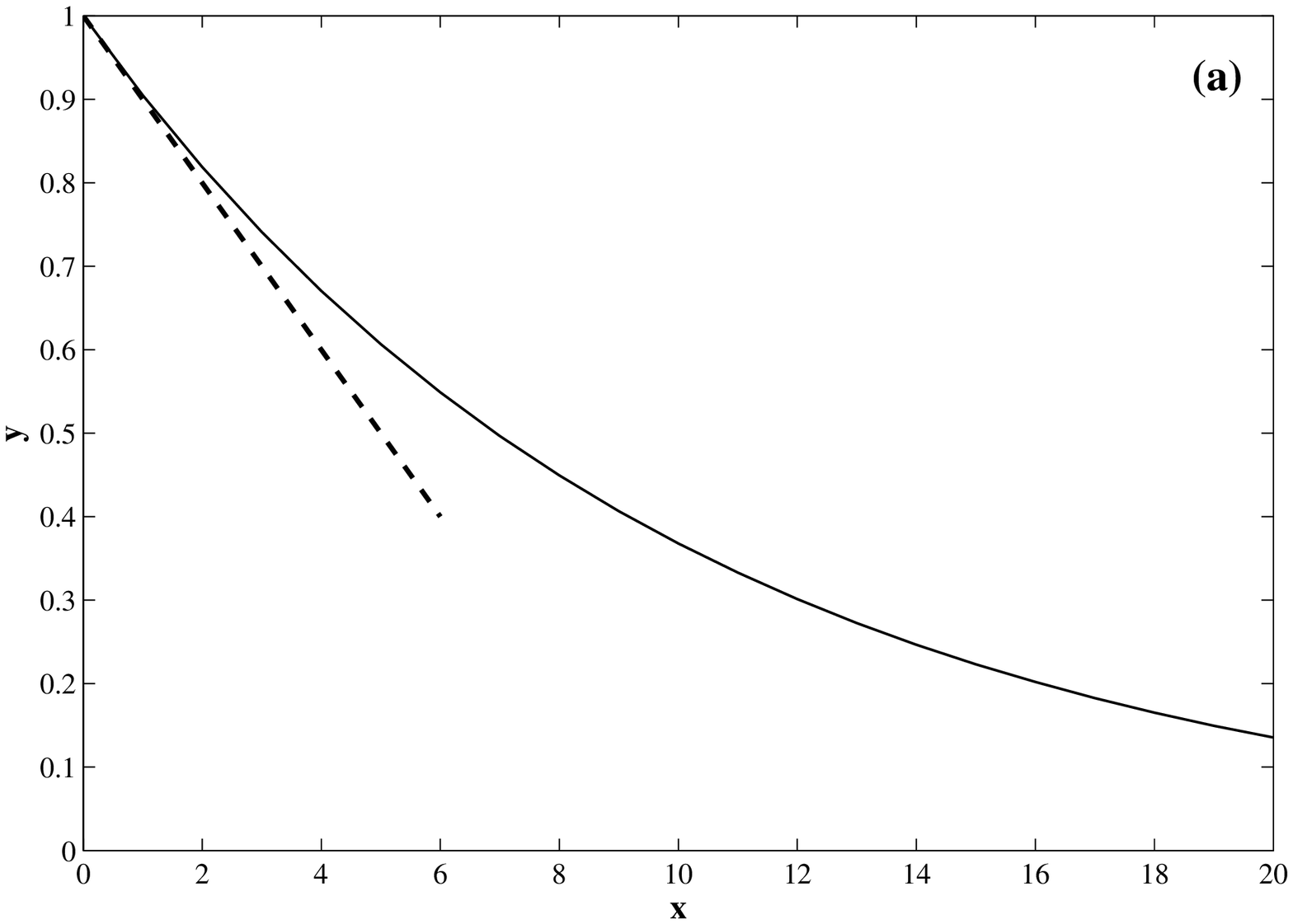}}
\vspace{0.1cm}
\hspace{-2.5cm}  
\resizebox{2\columnwidth}{!}{%
  \includegraphics{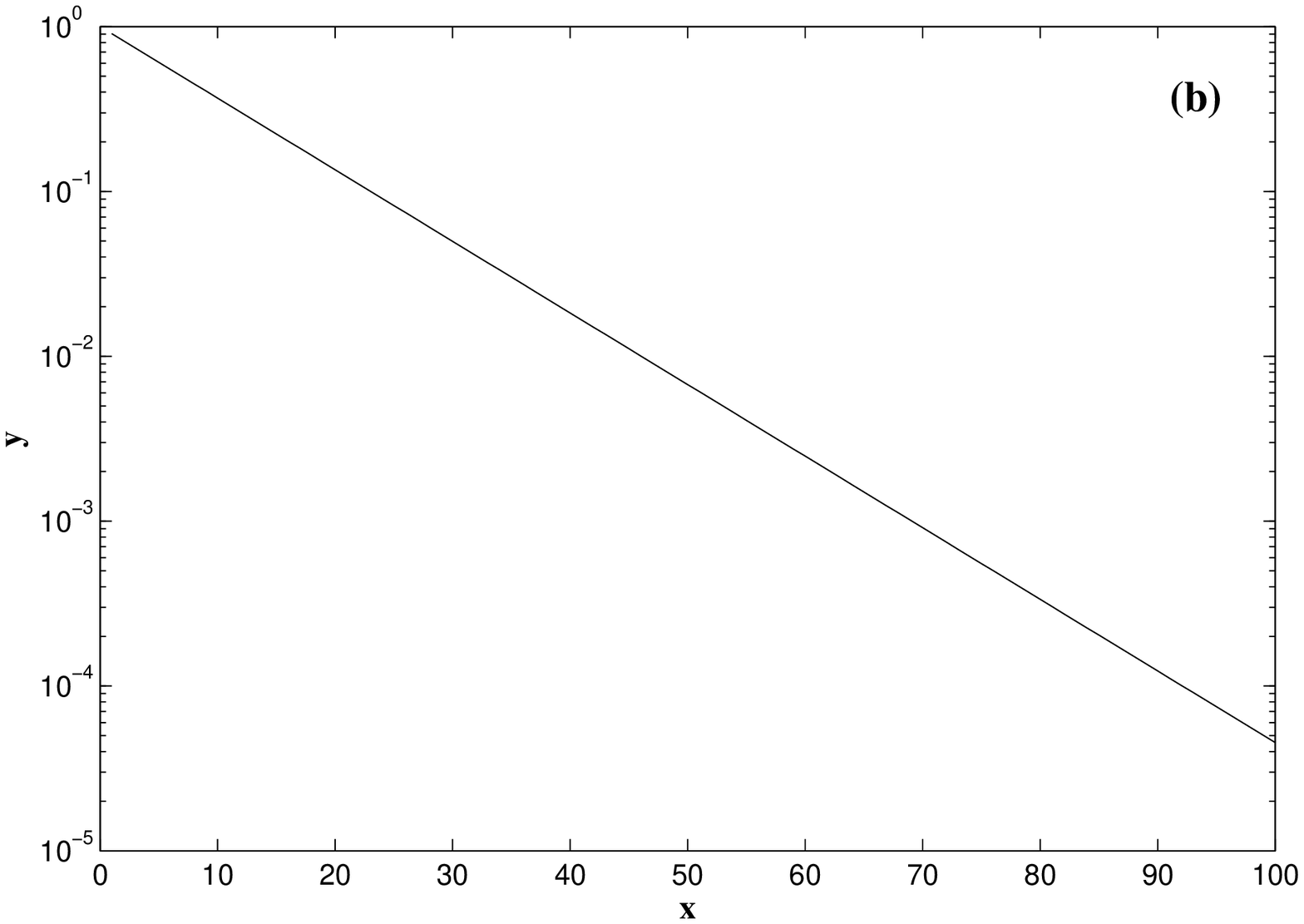}} 
\vspace{0.1cm}   
\hspace{-2.5cm}
\resizebox{2\columnwidth}{!}{%
  \includegraphics{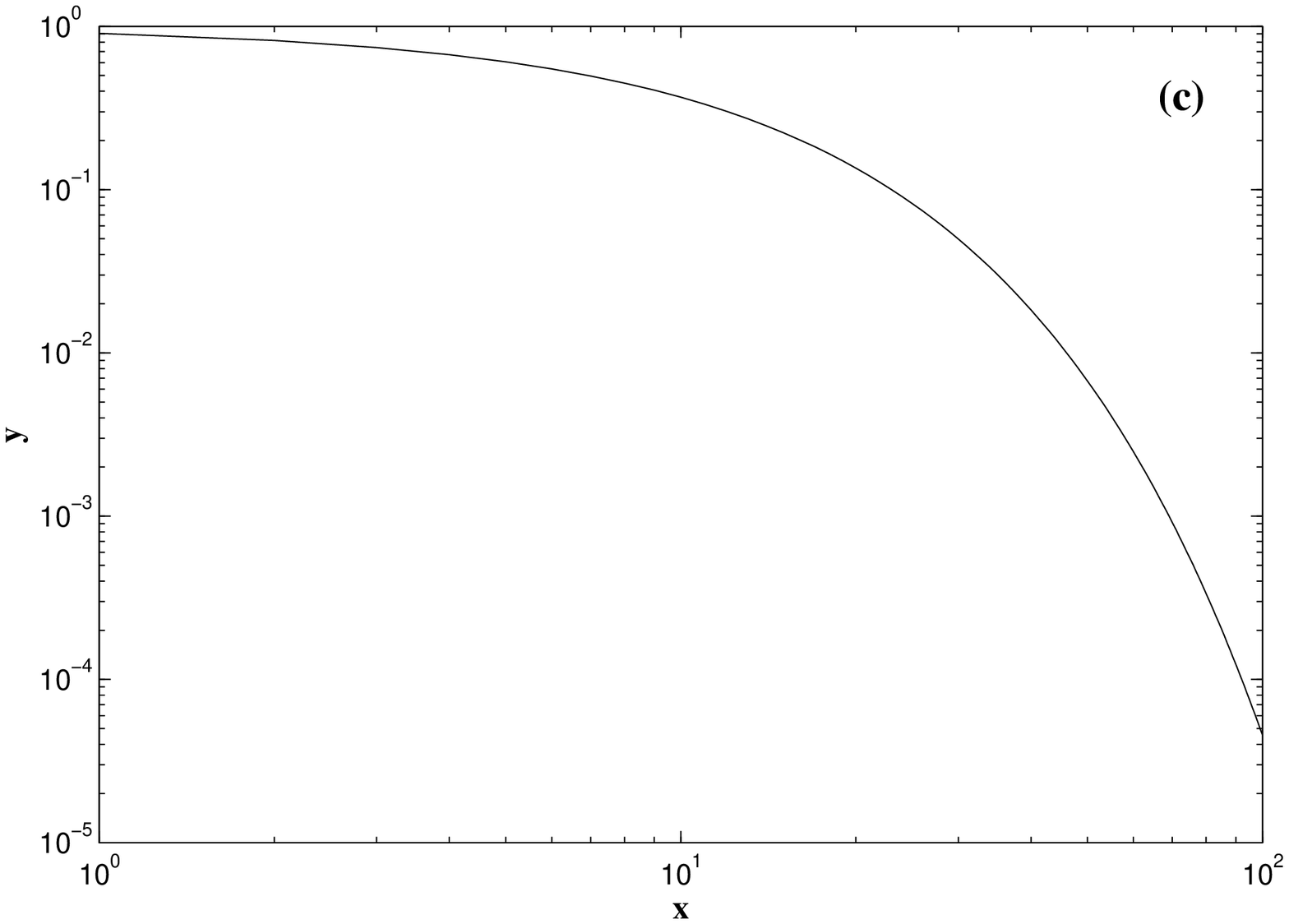}}   
\caption{The function $y=e^{-x/\xi}$ with $\xi =10$ in typical representations: (a) linear-linear; (b) log-linear; (c) log-log. In the linear-linear representation, the two regimes $x/\xi \to 0$ and $x/\xi \to \infty$ are clearly visible. Not so in the log-linear and log-log representations.}
\label{fig:1}    
\end{figure}

\section{Distinct physical mechanisms {\it versus} distinct regimes of a single physical mechanism}

We address in this section a question which emerges quite frequently in physics and elsewhere. It is in particular quite often present in the literature of experimental high energy physics. The question concerns within which context can/must we analyze the possible concurrence of various distinct physical mechanisms in a single set of experimental data or phenomenon. Let us first remind a historical illustration: electricity, magnetism and optics were discussed during centuries as being different physical mechanisms. But, since the celebrated unification performed in Maxwell equations, these phenomena are currently thought as {\it different asymptotic limits} (or {\it regimes}) of a {\it single mechanism}, rather than as the coexistence of {\it different physical mechanisms}. The same happened of course in the case of the weak and electromagnetic interactions. Therefore, along such discussions we must always have in mind that the opposition of different regimes against different physical mechanisms might, in this sense, be a historical and possibly provisory one.

Let us mathematically illustrate this in another famous example, namely Planck's law for the black-body radiation. It reads, for the photon energy density per unit volume,
\begin{equation}
u(\nu)=\frac{8 \pi h \nu^3}{c^3(e^{h \nu / k_B T}-1)} \,.
\end{equation}
This law presents a {\it low-frequency} asymptotic regime,
\begin{equation}
u(\nu)\sim 8 \pi  k_BT c^{-3}\nu^2 \,,
\end{equation}
as well as a {\it high-frequency} asymptotic regime
\begin{equation}
u(\nu)\sim 8 \pi h c^{-3}\nu^3e^{-h \nu / k_B T} \,,
\end{equation}
with an intermediate or crossover regime in between (where $u(\nu)$ is roughly constant). The high-frequency regime obviously is a quantum one (indeed, we see therein Planck constant $h$), whereas the low-frequency regime is a classical one. During many years these two regimes were considered as {\it two distinct physical mechanisms}. Since the formulation of quantum mechanics, however, many scientists prefer to think of them in a more ``modern" manner, namely as {\it two distinct asymptotic limits of a single physical mechanism}. 

In too many occasions, such discussions are regretfully driven, not by fundamental questions, but rather by the particular way of representing the results. For example, if we have the relation $y=e^{-x/\xi}$ with $x \ge 0$ and $\xi >0$, we can of course distinguish two asymptotic regimes, namely the $0 \le x <<\xi$ and the $x >> \xi$ ones. These two regions can be clearly identified in a {\it linear-linear} representation ($y \sim 1-(x/\xi)$ for $x / \xi \to 0$, and $y \sim 0$ for $x/\xi \to\infty$), less clearly however in a {\it log-linear} representation (where we only see a single straight line), or in a {\it log-log} representation (where we see a downwards ever bending curve): see Fig. 1. But, in such a situation, few scientists would keep talking of two distinct physical mechanisms; most of them would prefer to talk about two regimes of a single mechanism.  

\begin{figure}
\hspace{-2.5cm}
\resizebox{2\columnwidth}{!}{%
  \includegraphics{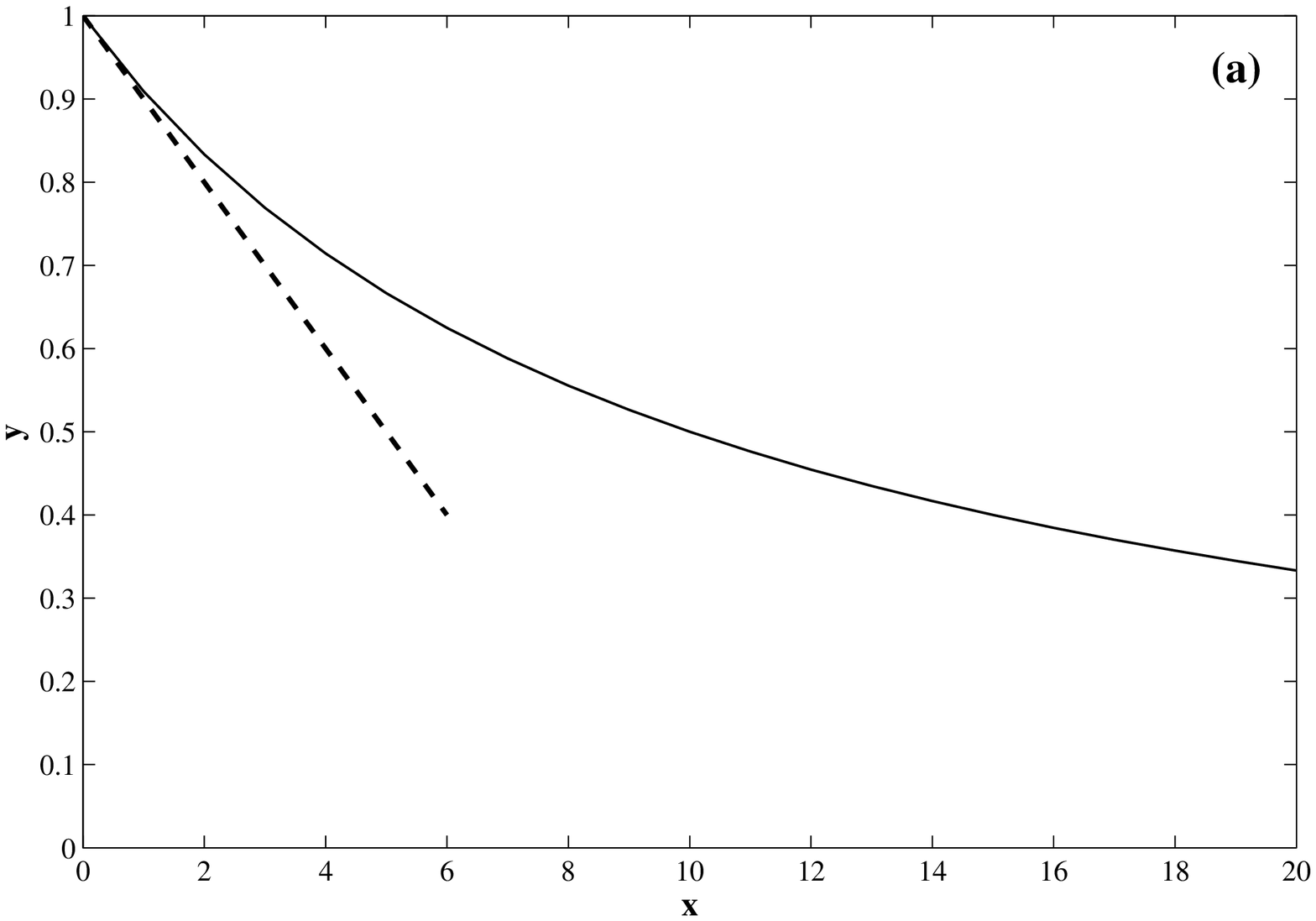}} 
\vspace{0.1cm}
\hspace{-2.5cm}  
\resizebox{2\columnwidth}{!}{%
  \includegraphics{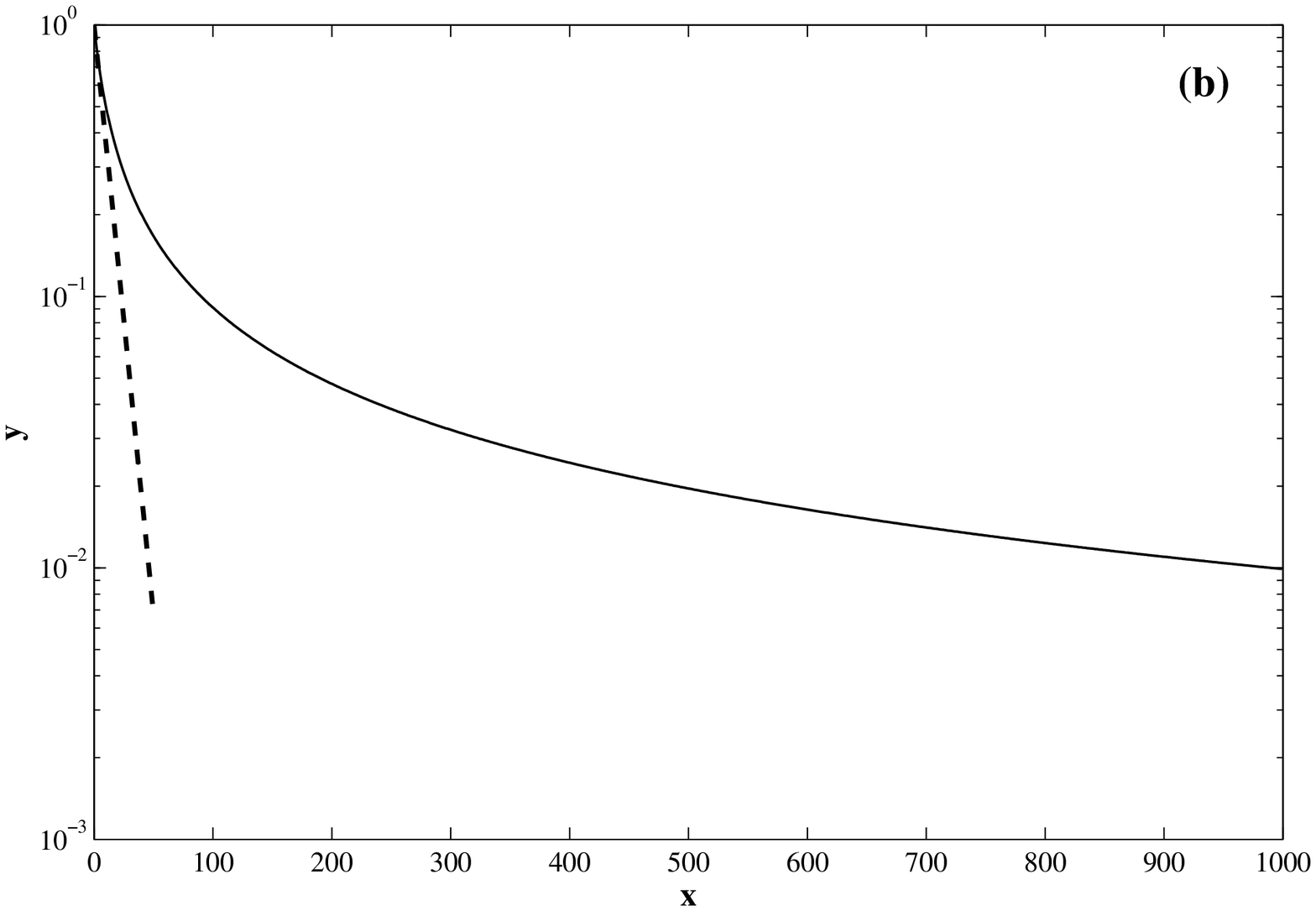}} 
\vspace{0.1cm}   
\hspace{-2.5cm}
\resizebox{2\columnwidth}{!}{%
  \includegraphics{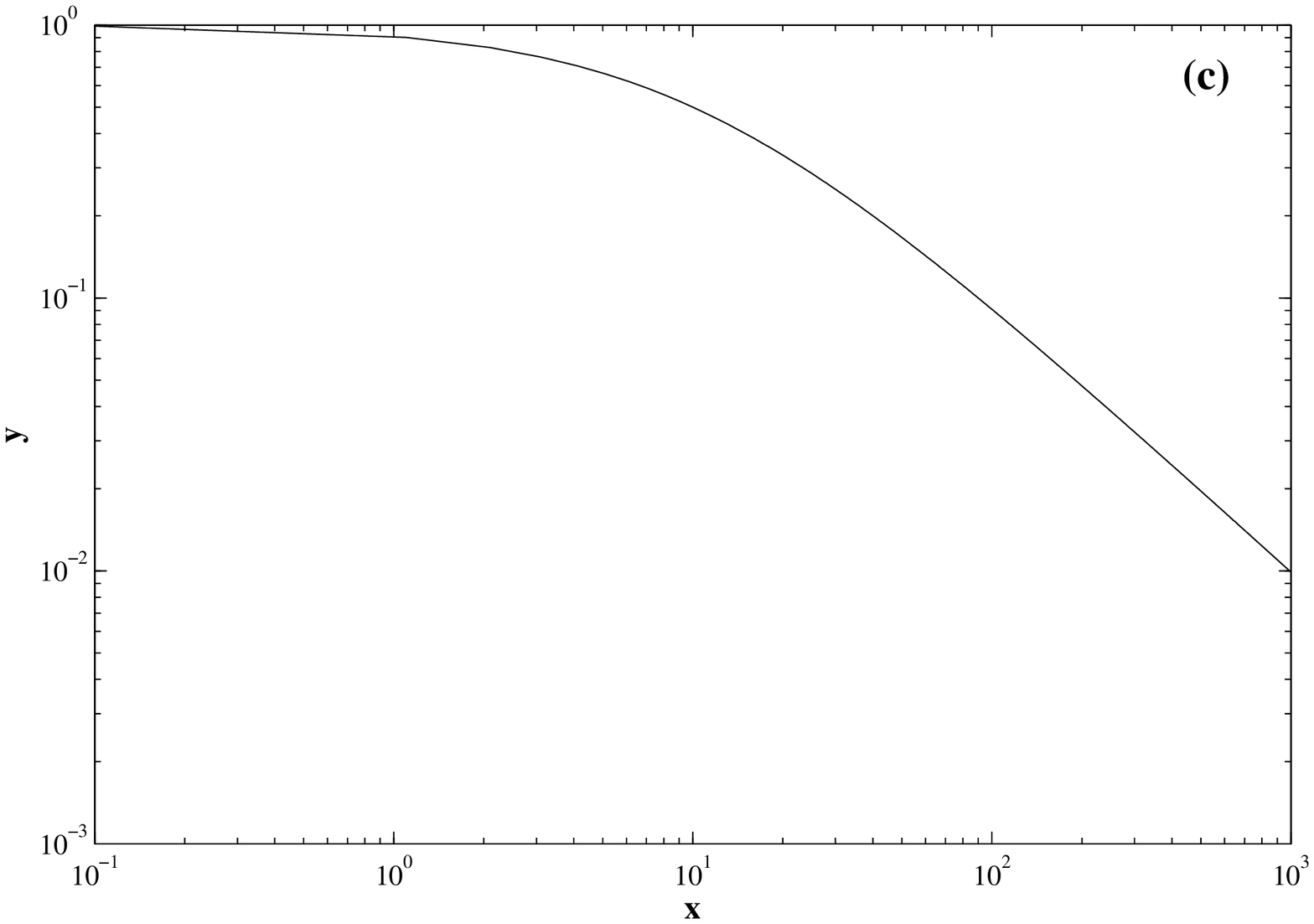}}     
\caption{The function $y=e_q^{-x/\xi}=1/[1+(q-1)x/\xi]^{1/(q-1)}$ with $q=2$ and $\xi =10$ in typical representations: (a) linear-linear; (b) log-linear; (c) log-log. 
In the linear-linear, log-linear (see the departure from a straight line) and log-log representations (see the roughly horizontal and the power-law regions), the two regimes $x/\xi \to 0$ and $x/\xi \to \infty$ are clearly visible. Not so in the (d) qlog-linear and (e) qlog-qlog representations, as can be seen in Fig. 3.}
\label{fig:2}    
\end{figure} 

Let us discuss now this very same point for a relation $y=e_q^{-x/\xi}=1/[1+(q-1)x/\xi]^{1/(q-1)}$ with $q>1$, which frequently emerges in high-energy collisions, and elsewhere. Once again, we can neatly distinguish two asymptotic regimes, namely the $x / \xi \to 0$  and the $x / \xi \to\infty$ ones. In the first regime we have $y \sim 1-(x/\xi)$ ($\forall q \ge 1$); in the second one, we have $y \sim 0$, or, in a more detailed form, $y \sim 1/[(q-1)x/\xi]^{1/(q-1)}$ ($\forall q>1$). Let us now check possible representations of this relation in Figs. 2 and 3. Like for the $q=1$ previous example, we clearly see the two regimes in the {\it linear-linear} representation. The novelty for $q>1$ is that we also see it in the {\it log-linear} and in the {\it log-log} representations. However, there is no a priori fundamental reason for now talking about two different mechanisms instead of just keep talking, as for the $q=1$ case, about two different asymptotic regimes. In fact, as can be seen in the $qlog-linear$ representation, a single straight line represents the whole curve, in total similarity with the $log-linear$ representation for the $q=1$ case. Also, analogously to the $q=1$ case, in a $qlog-qlog$ representation we do not clearly see the two regimes. These few simple remarks might help to put in a proper context the discussion of experimental data, thus avoiding naive conclusions and enabling a deeper understanding of the physical phenomena which are involved.

\section{Conclusions}

The $q$-exponential functions, frequently present (at least as approximations) in high-energy physics and elsewhere, emerge naturally within nonextensive statistical mechanics by optimizing the nonadditive entropy $S_q$ under appropriate constraints. Their origin at a mesoscopic level might include multiplicative noise, non-Markovian processes, space-dependent diffusion coefficients, nonlinear diffusion, and related mechanisms. The discussion at this level will naturally provide further understanding but will be in general insufficient for determining the precise value of  an index such as $q$. To achieve this deeper understanding, the analysis of a microscopic model (or, more precisely, a class of models) will be necessary. Typically, if the microscopic dynamics exhibits, for classical systems, vanishing Lyapunov exponents (or analogous situations for quantum systems), the system automatically becomes a strong candidate for the use of the nonextensive thermostatistical concepts. Such situations might occur, for instance, whenever the generic correlations are long-ranged in space-time (long-ranged interactions, quantum strongly entangled systems, etc). By studying quantities such as the sensitivity to the initial conditions as a function of time, one may obtain, from first principles, the value of the index $q$ (or of indices related to $q$).

\section*{Acknowledgments}
For the content of Section 3, I have benefited from an interesting conversation with E.M.F. Curado, I. Bediaga, J. Miranda and J. Takahashi, as well as from useful remarks by A.D. Anastasiadis. Partial support by Faperj and CNPq/MCT (Brazilian agencies) is acknowledged as well.

\begin{figure}
\hspace{-2.5cm}
\resizebox{2\columnwidth}{!}{%
  \includegraphics{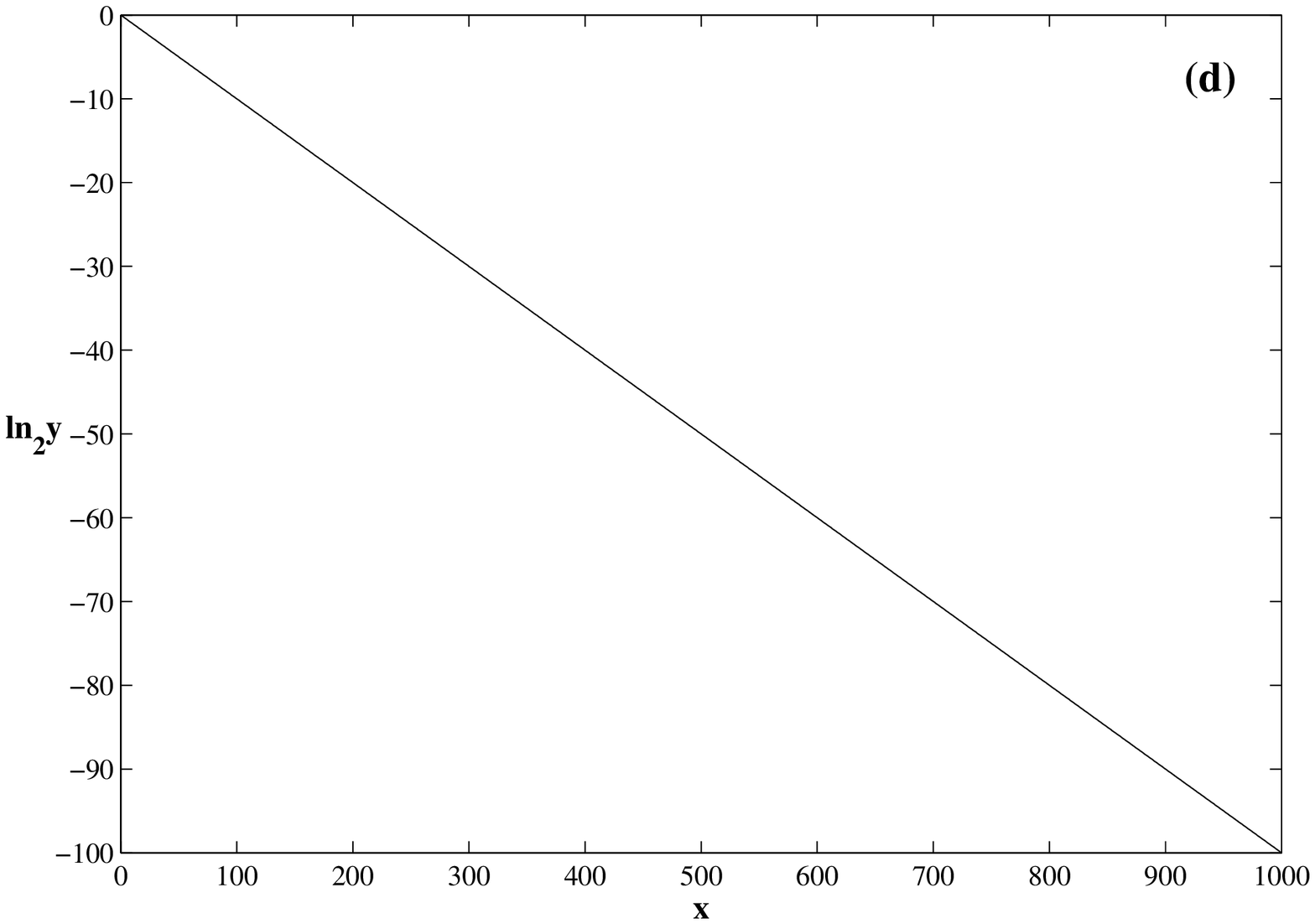}} 
\vspace{0.1cm}
\hspace{-2.5cm}  
\resizebox{2\columnwidth}{!}{%
  \includegraphics{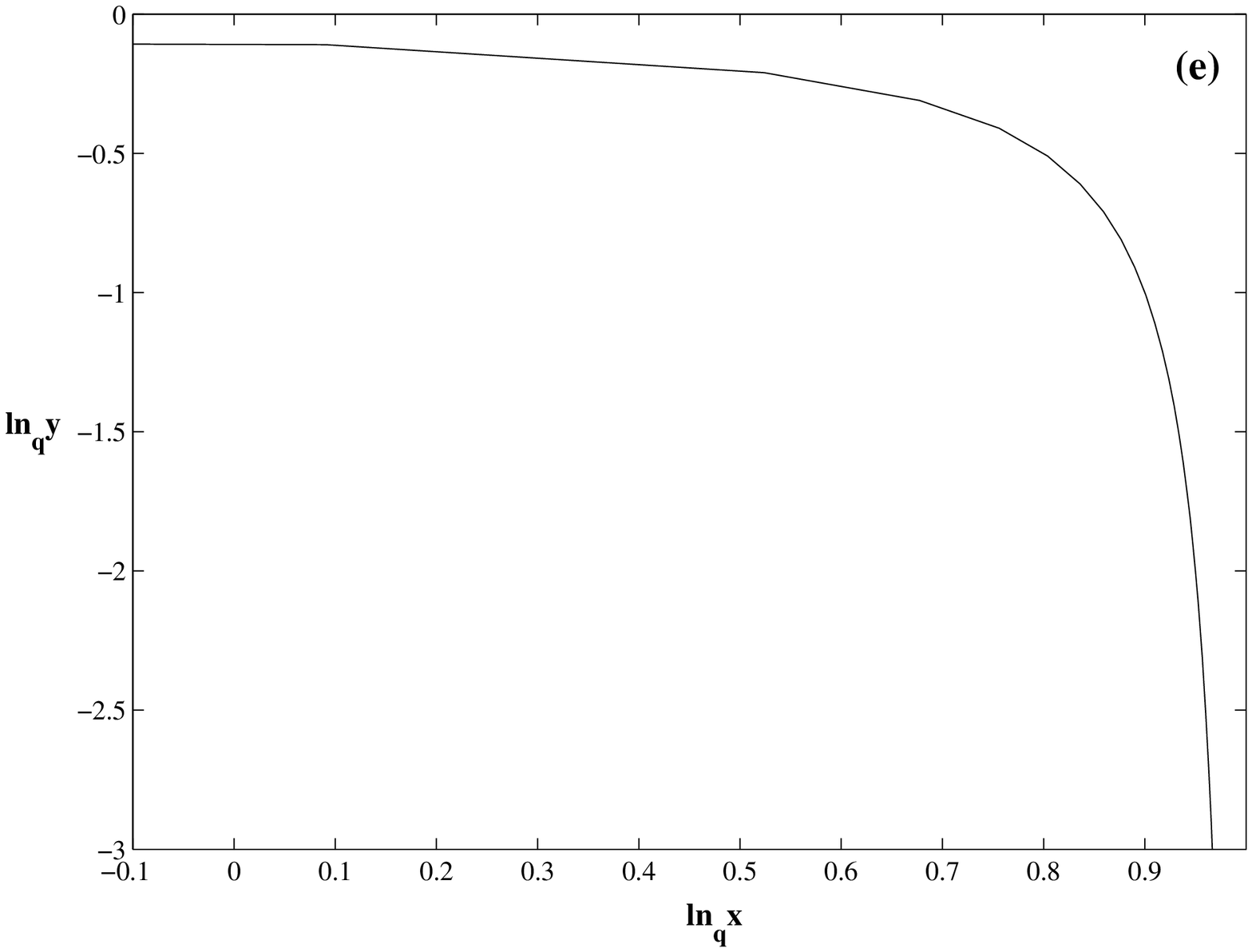}} 
\caption{The same as in Fig. 2. As a complement to the standard representations (a), (b) and (c) of Fig. 2, we indicate here the non-standard representations (d) $qlog-linear$, and (e) $qlog-qlog$. We remind that $\ln_q z \equiv \frac{z^{1-q}-1}{1-q}$ ($\ln_1 z=\ln z$). The two regimes $x/\xi \to 0$ and $x/\xi \to \infty$ are by no means clearly visible in these two unfamiliar representations.
}
\label{fig:3}    
\end{figure}

\end{document}